\documentclass[12pt]{iopart}

\usepackage{adjustbox}

\usepackage{booktabs}
\usepackage{multirow}
\usepackage{enumitem}
\usepackage{hyperref}

\usepackage{iopams}
\begin{document}
\title{Determinable and interpretable network representation for link prediction}

\author{Yue Deng}

\address{Institute of Fundamental and Frontier Sciences, University of Electronic Science and Technology of China,  Chengdu 611731, People’s Republic of China}
\ead{201921210214@std.uestc.edu.cn}
\vspace{10pt}

\begin{abstract}
As an intuitive description of complex physical, social, or brain systems, complex networks have fascinated scientists for decades. Recently, to abstract a network's structural and dynamical attributes for utilization, network representation has been one focus, mapping a network or its substructures (like nodes) into a low-dimensional vector space. Since the current methods are mostly based on machine learning, a black box of an input-output data fitting mechanism, generally the space's dimension is indeterminable and its elements are not interpreted. Although massive efforts to cope with this issue have included, for example, automated machine learning by computer scientists and computational theory by mathematics, the root causes still remain unresolved. Given that, from a physical perspective, this article proposes two determinable and interpretable node representation methods. To evaluate their effectiveness and generalization, this article further proposes Adaptive and Interpretable ProbS (AIProbS), a network-based model that can utilize node representations for link prediction. Experimental results showed that the AIProbS can reach state-of-the-art precision beyond baseline models, and by and large it can make a good trade-off with machine learning-based models on precision, determinacy, and interpretability, indicating that physical methods could also play a large role in the study of network representation.
\end{abstract}

%
%
%
%
%

\section{Introduction}

Physics has long been concerned as a propeller of civilization's evolution in history. The establishment of Newtonian mechanics and thermodynamics drove the ``first technological revolution''. The discovery of the electromagnetic induction phenomenon laid the theoretical foundation for the ``second technological revolution''. Condensed matter physics and quantum physics developed the silicon semiconductor industry for the ``third technological revolution''. With the ongoing ``fourth technological revolution'' currently, physics is also propelling the innovation and development in artificial intelligence, among which the study of complex networks \cite{strogatz2001exploring, boccaletti2006complex} is a case in point. By using nodes and edges to intuitively describe the nonlinear and heterogeneous interaction patterns of components composing the complex physical, social, or brain systems, its use soon widened to various fields. For decades, scientists have been dedicated to understanding a network's structural and dynamical attributes (like vital node identification \cite{zhou2019fast, qiu2021identifying}, high-order network structural analysis \cite{shi2019totally, battiston2020networks, shi2021computing,tang2022optimizing}, and percolation theory \cite{li2021percolation}) and to utilizing these attributes in specific applications, such as link prediction \cite{lu2011link}, natural language processing \cite{pardo2006using}, and recommender systems \cite{lu2012recommender, xu2020recommending, deng2021recommender}.

Recently, as a pivotal tool to abstract a network's structural and dynamical attributes for utilization in a manner that maps the network or its substructures (like nodes) into a low-dimensional vector space, network representation \cite{liu2021network, barros2021survey} has intrigued scientists for years, especially in light of ample evidence that network representation has several virtues dear to both academia and industry \cite{deng2021recommender}: reusable object representations by manual or automated feature engineering, enhanced model precision, and efficient parallel computation based on GPU, among which some are even unprecedented compared to their predecessors. Nevertheless, since the current methods of network representation are mostly based on machine learning \cite{shalev2014understanding}, almost a black box facing fundamental limits on well-explainable and raising difficulties in tedious hyperparameter tuning attributed to its input-output data fitting rationale, the learned vector space's dimension is generally indeterminable and its elements are not interpreted. Consequently, enormous computing resources are required for searching the suboptimal dimension for the vector space within a range but in most cases researchers still can not interpret why such a dimension works out and what realistic meanings the space's elements may represent. Although recent years have seen massive efforts by computer scientists and mathematics to cope with this issue, the root causes still remain unresolved. For example, although automated machine learning \cite{yao2018taking} proposed by computer scientists can pave the way for accelerating the search of the space's suboptimal dimension not manually, the search mechanism requiring enormous computing resources still remains. Moreover, although mathematics can conclude an empirical formula used to select the optimal dimension for the space \cite{gu2021principled}, it is normally built on specific models or data, limiting its interpretability and generalization to other scenarios. Given these inadequacies, determinable and interpretable network representation is still an open and important question.

In this work, from a physical perspective, this article proposes two methods of determinable and interpretable network representation. Methodologically, the first method is based on the degree, H-index, and Coreness (DHC) theorem \cite{lu2016h} constructing an operator to generate sequences (with a fixed length) of H-indices for nodes. Regarding their realistic meanings and on the advice of the rich club theory \cite{colizza2006detecting}, this article utilizes these H-indices to construct node representations, which can represent nodes' local attributes around the neighborhood. To abstract nodes' global attributes in a complex network, the second method is based on the DHC entropy (DHC-E) \cite{wang2021hyperparameter}, a hyperparameter-free and explainable whole graph embedding algorithm we proposed. If a bipartite network, its $m \times n$ adjacent matrix can be extended to a $(m+n) \times (m+n)$ augmented matrix, a simple matrix that can be decomposed to $m+n$ matrices, each of which corresponds to a node and carries the node's global attributes. After implementing the DHC-E algorithm on them, node representations can be generated. Unlike those learned by machine learning-based methods, the node representations generated by the two methods have both a determined dimension and interpretable elements.

To evaluate the proposed two methods' effectiveness and generalization, this article further proposes Adaptive and Interpretable ProbS (AIProbS), a network-based link prediction model for bipartite networks, which can utilize nodal representations generated by the two methods, as an attempt to enhance the prediction precision. Methodologically, built on a classical network-based framework called ProbS \cite{zhang2007recommendation}, the AIProbS can control the resource diffusion process of the ProbS framework by setting edge weights quantified with node representations, which can perceive the similarity between nodes. After being equipped with such artificial intelligence as machine learning-based models do, the AIProbS makes the flaw of the classical ProbS framework in self-adaptive perception ability oriented to different scenarios (which is analyzed in Sec.~\ref{ProbS}). At the same time, compared with machine learning-based link prediction models \cite{koren2008factorization, rendle2012bpr, chen2019integrating, jiang2020clustering, xu2021topic}, the AIProbS is hyperparameter-free. In addition, implemented on several designed control experiments of diverse recommender systems (a specific application of link prediction in artificial intelligence), experimental results showed that the AIProbS can reach state-of-the-art precision beyond baseline models on some scenarios and can, by and large, make a good trade-off with machine learning-based models on precision, determinacy, and interpretability.

\section{The model}

In the first place, this article proposes two novel network representation methods in Sec.~\ref{generate_node_features}. Then, a classical network-based link prediction framework called ProbS is introduced and its flaws are revealed in Sec.~\ref{ProbS}. Based on the ProbS framework, this article proposes Adaptive and Interpretable ProbS (AIProbS) in Sec.~\ref{AIProbS}, a network-based link prediction model for bipartite networks, which can utilize nodal representations generated by the two methods and can enhance the prediction precision of the classical ProbS framework by making up its flaws.

\subsection{Generate a complex network's nodal representations}
\label{generate_node_features}

\subsubsection{Method one}
\label{method_one}

Degree, H-index, and coreness are three measurements used to quantify nodal influence in a complex network. Node's degree measures nodal influence by counting a node's neighbors: the greater a node's degree is, the more neighbors it is connected with, and the higher influence it has. Node's H-index \cite{hirsch2005index} is the maximum value $h$ such that a node has at least $h$ neighbors with a degree no less than $h$. Furthermore, to take location into account, coreness calculated by $k$-core decomposition \cite{dorogovtsev2006k} measures a node's centrality: a greater coreness indicates that a node locates more centrally in a complex network and hence has a higher influence.

The DHC theorem \cite{lu2016h} reveals that degree, H-index, and coreness are all related. To describe the relationship, the DHC theorem constructs an operator $\mathcal{H}$, which calculates the maximum value $h$ for each node such that the node has at least $h$ neighbors with H-indices no less than $h$. For each node $i$ in a complex network, taking its degree $k_i$ as the zero-order H-index $h_i^{(0)}$ as the beginning, the first-order H-index $h_{i}^{(1)}$ of node $i$ is calculated by $\mathcal{H}(h_{j_1}^{(0)}, h_{j_2}^{(0)}, ..., h_{j_{k_i}}^{(0)})$, where $h_{j_1}^{(0)}, h_{j_2}^{(0)}, ..., h_{j_{k_i}}^{(0)}$ are the zero-order H-indices (\textit{i.e.}, the degree values) of the $k_i$ neighbors of node $i$. By iteratively doing so, $h_i^{(2)} = \mathcal{H}(h_{j_1}^{(1)}, h_{j_2}^{(1)}, ..., h_{j_{k_i}}^{(1)})$, as well as $h_i^{(3)}, h_i^{(4)},...$, can be calculated. Finally, a sequence $h_i^{(0)}, h_i^{(1)}, h_i^{(2)}, ...$ with a fixed length is generated for node $i$, which is convergent to node $i$'s coreness, as the DHC theorem states:

\textbf{Theorem 2.1}. \textit{For each node in a complex network, node $i$'s H-indices sequence $h_i^{(0)}, h_i^{(1)}, h_i^{(2)}, ...$ is convergent to its coreness $c_i$, \textit{i.e.}, $ \displaystyle c_i = \lim_{n \to \infty} h_i^{(n)}$.}

\textit{Proof.} See \cite{lu2016h}.

According to the rich club theory \cite{colizza2006detecting} (from the field of social network analysis \cite{scott1988social} and soon widened to interdisciplinary studies like computer science \cite{zhou2004rich} or cognitive science \cite{van2011rich}) that a node's influence could reflect its attributes and functions around the neighborhood and the whole network structure, this article proposes the following assumption:

\textbf{Assumption 2.1}. \textit{A node's H-indices sequence can abstract the node's multidimensional influence in the neighborhood, where the sequence's convergence steps can reflect the magnitude of the node's influence. The more important role played by the node in the neighborhood, the more slowly its influence decays during the dynamic evolution (\textit{i.e.}, the convergence process by the DHC theorem), thus the larger its convergence steps are. }

Built on assumption 2.1 this article takes a node's H-indices sequence as its node representation. In this way, provided $n$ nodes in a complex network and given that their H-indices sequence converges after up to $s$ steps, this method can map the $n$ nodes to a $s$-dimensional vector space consisting of their H-indices as node representations. This is a determinable and interpretable network representation method, since for an arbitrary complex network the dimension of its nodal representations is determined as $s$ and the elements can be interpreted as nodal multidimensional influence with different magnitudes.

\subsubsection{Method two}
\label{method_two}

Following method one, to further abstract a node's global attributes in a complex network, if a bipartite network, its adjacency matrix $A^{m \times n}$ can be extended to $B^{(m+n)\times (m+n)}$ constructed by
$\displaystyle \left(\begin{array}{cc}
O^{m \times m}        & A^{m \times n} \\
(A^{m \times n})^T    & O^{n \times n}
\end{array}\right)$,
where $O$ denotes the null matrix. Based on it, a series of $\lambda_i$ and $B_i$ can be decomposed by the following theorem.

\textbf{Theorem 2.2}. \textit{The adjacency matrix $\displaystyle B^{(m+n)\times (m+n)}$ can be decomposed by $\displaystyle B=\sum_{i=1}^{m+n}\lambda_i B_i$, where $\lambda_i$ is the i-th eigenvalue of $\displaystyle B^{(m+n)\times (m+n)}$ and $\displaystyle B_i$ is the corresponding idempotent matrix.}

\textit{Proof.} See \textbf{\textit{Appendix A}}.

After that, this article implements the DHC-E operator $\mathcal{E}$ \cite{wang2021hyperparameter} (\textit{i.e.}, by the DHC theorem to generate a H-index matrix $H^{n \times s}$ by row containing the H-indices converged after $s$ steps of each of the $n$ nodes in a complex network, the operator $\mathcal{E}$ calculates the Shannon entropy of each column of $H^{n \times s}$ and obtains a vector $e^{1 \times s}$, as the whole graph embedding of the network) on each $B_i$ or $\lambda_i B_i$ one by one, generating the $m+n$ nodes' representations for the bipartite network correspondingly. Apparently, this method is also a determinable and interpretable network representation method. The characteristics of interpretability and hyperparameter-free of the DHC-E algorithm are thoroughly illuminated in \cite{wang2021hyperparameter}.

\subsection{The ProbS framework and its flaws}
\label{ProbS}

To evaluate the two methods' effectiveness and generalization, this article utilizes them in link prediction for bipartite networks. Since network representation can be seen as artificial intelligence that recognizes and abstracts a complex network's underlying structural and dynamical attributes, this article explores how such artificial intelligence (\textit{i.e.}, nodal representations generated by these methods) can be used to enhance the precision of classical link prediction models.

Among classical (non-machine learning-based) link prediction models for bipartite networks, the ProbS \cite{zhang2007recommendation} framework is a typical one. By means of a resource diffusion mechanism inspired by the physical process of Material Diffusion, the ProbS framework can quantify the similarity between nodes after initializing and diffusing resources. Fig.~\ref{Schematics} includes an example to intuitively illuminate the schematics of the ProbS framework. For instance, when predicting node $B$'s unobserved links with nodes $a$ and $b$, resources are first initialized at nodes $c$ and $d$ (the nodes that are connected with node $B$) with value $1$, then are diffused to nodes $A$, $B$, and $C$ along edges after being equally divided by the degree of each node, finally are diffused back to nodes $a$, $b$, $c$, and $d$ in the same way, which can be used to quantify the similarity between node $B$ and the four nodes, respectively. A larger similarity of two nodes indicates a higher probability of an unobserved link existing between them.

This article provides a mathematical perspective to describe the ProbS framework, by constructing an operator $T$ to describe its diffusion mechanism. Given a bipartite network consisting of $m+n$ nodes of two different types, respectively, whose adjacency matrix is represented by $A^{m \times n}$. Let $R^{m \times n}$ denote the predicted matrix, where $R_{ij}$ represents the similarity (\textit{i.e.}, the probability of the existence of a link) between nodes $i$ and $j$. Then, through the ProbS framework $R^{m \times n}$ can be calculated by
\begin{equation}
\label{Probs}
    R = A \cdot (D_{I}  \circ  A)^T \cdot (D_{U}  \circ  A)
\end{equation}
where $\cdot$ denotes the dot product, and $\circ$ denotes the Hadamard product. $D_I^{m \times n} = (a_1, a_2, ..., a_n)$, $\displaystyle a_i = (\frac{1}{k_{I_i}}, ... , \frac{1}{k_{I_i}})^T$ where $k_{I_i}$ is item $i$'s degree. $D_U^{m \times n} = (a_1, a_2, ..., a_m)^T$, $\displaystyle a_i = (\frac{1}{k_{U_i}}, ... ,\frac{1}{k_{U_i}})$ where $k_{U_i}$ is user $i$'s degree. In Eq.~(\ref{Probs}) the operator $T=(D_{I}  \circ  A)^T \cdot (D_{U}  \circ  A)$.

The operator $T$ tells why the ProbS framework will converge after deriving $R$ from $A$ and then placing $A$ with the derived $R$ iteratively, stated as the following theorem.

\textbf{Theorem 2.3}. Let the operator $T=(D_{I}  \circ  A)^T \cdot (D_{U}  \circ  A)$ iteratively act on $A$ by $A \leftarrow A \cdot T$, the iterative process is convergent.

\textit{Proof}. See \textbf{\textit{Appendix B}}.

Since the difference between the values in $A$ tends to be smoother as the convergent iterative process progresses but link prediction relies for higher precision on the more distinctive differentiation between the predicted values of similarity \cite{deng2021recommender}, in link prediction the best iteration steps for the ProbS framework is one, and so does the AIProbS proposed in Sec.~\ref{AIProbS}.

In addition, from such a mathematical perspective, it is easy to see that the ProbS framework faces fundamental limits on intelligence because its resource diffusion mechanism is just based on equal allocation, shown as $D_I$ and $D_U$ in Eq.~(\ref{Probs}). In practice like recommender systems (an application of link prediction for bipartite networks in artificial intelligence), such a mechanism raises a key question: if respectively take these nodes of two different types as users and items in recommender systems, the resources diffused between users and items back and forth, to some extent, represent user's preferences for items or item's attractiveness to users, while neither of them should be necessarily equal since user biases \cite{koren2009matrix, adomavicius2014biasing, manjur2021exploring} and item biases  \cite{koren2009matrix, park2014uncovering} generally exist in reality. Moreover, these biases are usually recommendation scenario-oriented, which means that in different scenarios a user's preferences may differ, and so do an item's attractiveness or popularity. Finally, in practice the ProbS framework fails to take these biases into consideration, let alone adaptively perceive and quantify their differences in various scenarios.

\subsection{The AIProbS model}
\label{AIProbS}

The essential condition for the ProbS framework to realize that intelligence is to be equipped with self-adaptive perception, an ability to perceive and utilize the attributes of nodes (\textit{i.e.}, nodal representations) in a complex network toward different scenarios. To utilize the nodal representations generated by the two methods proposed in Sec.~\ref{generate_node_features} in the ProbS framework, this article proposes Adaptive and Interpretable ProbS (AIProbS).

In the first step, on the advice that the rich club theory \cite{colizza2006detecting} gives clues that nodes with high centrality tend to form tightly interconnected communities, this article generalizes this conclusion to the field of link prediction, proposing the following assumption:

\textbf{Assumption 2.2}. \textit{The similarity between node pairs having strongly correlated nodal representations (\textit{i.e.}, similar features or similar influence) is higher than that between weakly correlated ones.}

To measure the similarity between nodes, the AIProbS uses the cosine similarity metric. Provided two $n$-dimension vectors $x$ and $y$, the cosine similarity between them is calculated by $\displaystyle \cos(\theta)=\frac{x \cdot y}{|x| \cdot |y|}=\frac{\sum_{i=1}^n x_i y_i}{\sqrt{\sum_{i=1}^n x_i^2} \cdot \sqrt{\sum_{i=1}^n y_i^2}}$. In the same way, provided $m+n$ nodes belonging to two sets $U$ and $I$ of two different types in a bipartite network, respective. Through the network representation methods proposed in Sec.~\ref{generate_node_features} the representation matrices $F_U^{m \times s}$ and $F_I^{n \times s}$ of the two types of nodes are generated, either of which is consist of nodal representations by row. Then, the $m \times n$ nodal similarity matrix $S^{m \times n}$ calculated by the cosine similarity metric is
\begin{equation}
\label{user-item_similarity_matrix}
S = \frac{F_{U} \cdot F_{I}^T}{\alpha^T \cdot \beta},
\end{equation}
where vector $\displaystyle \alpha = \big(\sqrt{\sum_{j=1}^s{F_U}_{1j}^2}\,,\sqrt{\sum_{j=1}^s{F_U}_{2j}^2}\,, \,...\,, \sqrt{\sum_{j=1}^s{F_U}_{mj}^2}\big)$ and vector $\displaystyle \beta = \big(\sqrt{\sum_{j=1}^s{F_I}_{1j}^2}\,,\sqrt{\sum_{j=1}^s{F_I}_{2j}^2}\,, \,...\,, \sqrt{\sum_{j=1}^s{F_I}_{nj}^2}\big)$.

After obtaining the nodal similarity matrix $S^{m \times n}$, utilizing it to control the diffusion process of the classical ProbS framework is the second step. To assign proper weights to every node pair for the diffusion mechanism of the ProbS framework, the AIProbS further complete some normalization and proportioning operations on $S \circ A$ where $A$ is the adjacency matrix shown in Eq.~(\ref{Probs}). Since the elements of $S \circ A$ vary in $[-1,1]$ while the diffused resources are supposed to be positive, the AIProbS normalizes the value range of the elements to $[0,1]$ using the max-min normalization operation, for each row vector $(S\circ A)_{i*}$ $(i=1,2,...,m)$ of $S \circ A$, by
\begin{equation}
\label{max-min_normalization}
(S \circ A)_{ij} \leftarrow \frac{(S \circ A)_{ij} - \min}{ \max - \min}, \,j = 1,2, ..., n,
\end{equation}
where the $\max$ and $\min$ are the maximum and minimum elements of the row vector $(S \circ A)_{i*}$, respectively. Based on that, the weight matrix $W_U$ for nodes belonging to set $U$ is calculated by the proportioning operation as
\begin{equation}
\label{proportioning}
W_{U_{ij}} = \frac{1}{(S \circ A)_{ij}}\sum_{k=1}^{n} (S \circ A)_{ik}, \,i=1,2,...,m, \, j=1,2,...,n.
\end{equation}
On the other hand, the same operations are completed on $S^{m \times n}$ by column, generating the weight matrix $W_I^{m \times n}$ for nodes belonging to set $I$.

In the last step, the predicted matrix $R^{m \times n}$, where $R_{ij}$ represents the prediced similarity between nodes $i$ and $j$, is calculated through the AIProbS by
\begin{equation}
    R = A \cdot W_{I}^T \cdot W_{U}.
\end{equation}

Conceivably, there are other metrics for similarity measurement. More combinations were tested in this article (See \textbf{\textit{Appendix C}} for details) but none of them performed better than the one proposed in this section. All in all, the whole process of the AIProbS are summarized in the pseudocodes shown in \textit{Appendix D}. For more intuitive illumination, Fig.~\ref{Schematics} in \textbf{\textit{Appendix D}} presents its schematics.

\subsection{Performance Evaluation}
\label{performance_evaluation}

To evaluate the AIProbS's precision as well as its pros and cons in link prediction for bipartite networks, which also can be used to reflect the effectiveness of nodal representations generated by the two network representation methods proposed in Sec.~\ref{generate_node_features}, this article designs control experiments based on recommender systems, an application of link prediction in artificial intelligence.

\subsubsection{Recommender systems}

By analyzing observed user-item relations to predict a user's preferred items from millions of candidates, recommender systems \cite{lu2012recommender, deng2021recommender} are recognized as a pivotal tool to alleviate the information overload problem. Among different user-item relations, implicit user-item interactions (\textit{e.g.}, user's historical clicks or buys on items) record the existence of a user's interactions with items, defined as a binary state using $1$ and $0$. From the perspective of a complex network, recommendation on implicit user-item interactions can be seen as a process of link prediction for bipartite networks, where users and items correspond to the two types of nodes and implicit user-item interactions represent the edges between nodes. Therefore, the designed experiments in this article are based on the recommendation with implicit user-item interactions, for most current models are based on them.

\subsubsection{Data sets}

In light of the no-free-lunch theorem \cite{adam2019no} that no model can always perform well enough as expected in all different scenarios, this article designs control experiments to evaluate the performance of the AIProbS on diverse real recommendation scenarios, in order to explore not only the AIPobS's pros but also its cons in different scenarios.

\begin{table}[ht]
\caption{\textbf{Overview of data sets.}}
\label{Datasets}
\begin{indented}
\item[]\begin{tabular}{ccllcccc}
\br
Data sets       & \multicolumn{3}{c}{$|U|$} & $|I|$ &  Interactions & Sparsity  \\
\mr
MovieLens 100K & \multicolumn{3}{c}{943}            & 1680           & 100000           & 93.70\%                              \\
MovieLens 1M   & \multicolumn{3}{c}{6040}           & 3952           & 1000209          & 95.81\%                           \\
LastFM    & \multicolumn{3}{c}{1892}          & 17632            & 92834           & 99.72\%                            \\
\br
\end{tabular}
\end{indented}
\end{table}

As shown in Tab.~\ref{Datasets}, $|U|$ and $|I|$ represent the number of users and items, respectively, and the interactions between users and items are implicit ones. The sparsity in Tab.~\ref{Datasets} represents the ratio of the number of unobserved interactions to the maximum number of all possible interactions between users and items (e.g., that between $m$ users and $n$ items is $mn$). As a control group, the MovieLens 100K, MovieLens 1M, and LastFM are three classical data sets from two different recommender systems of movies and music, with distinctive ratios of $|U|$ to $|V|$, data scales, and sparsity, based on which more persuadable results could yield compared with those based on newly published data sets, since these classical data sets have been widely used for evaluation in previous works

In order to guarantee the reproducibility of experiments, either of the three data sets is obtained from the RecBole public resources (\url{https://recbole.io/dataset_list.html}), organized into tuples (user, item, $0/1$) without preprocessing. Each of them is randomly split into a ``train/evaluate/test'' set by the ratio of ``$80/10/10\%$''. After independently repeating the splitting process $30$ times, $30$ realizations are generated for each data set. One can get the split data used in this article through the hyperlink address posted in Sec.~\ref{data_and_code_available}.

\subsubsection{Evaluation metrics}

In order to quantify the precision of the AIProbS on these data sets, three common-used metrics are chosen in this article. Given a user $u \in U$ ($U$ is the user set) and the length $N$ of the recommendation list, the set of recommended items for the user is denoted by $\hat{R}(u)$ and the ground-truth set of items the user interacted with is denoted by $R(u)$. Based on them, the first evaluation metric is the Recall@N \cite{olson2008advanced}, which calculates the fraction of predicted relevant items out of all ground-truth relevant items by
\begin{equation}
\mbox{Recall@N}=\frac{1}{|U|}\sum_{u \in U} \frac{|\hat{R}(u) \cap R(u)|}{|R(u)|},
\end{equation}
where $|R(u)|$ represents the item count of $R(u)$.

To calculate the reciprocal rank of the first relevant item recommended to each user, the second evaluation metric MRR@N \cite{craswell2009mean} is denoted as
\begin{equation}
\mbox{MRR@N}=\frac{1}{|U|} \sum_{u \in U} \frac{1}{\mbox{rank}_u^*},
\end{equation}
where $\mbox{rank}_u^*$ is the rank position of the first relevant item recommended to user $u$.

Moreover, as the third evaluation metric, the NDCG@N \cite{wang2013theoretical} can further measure the overall ranking quality in a manner that accounts for the position of the hit by assigning higher scores to hits at top ranks as
\begin{equation}
\footnotesize
\mbox{NDCG@N}=\frac{1}{|U|} \sum_{u \in U} \big( \frac{1}{\sum_{i=1}^{\min(|R(u)|,N)}\frac{1}{\log_2(i+1)}}  \sum_{i=1}^N \delta(i \in R(u)) \frac{1}{\log_2(i+1)}\big),
\end{equation}
where $\delta(\cdot)$ is an indicator function and positions are discounted logarithmically.

In practice, the greater the values of these evaluation metrics are, the higher a model's precision is.

\subsubsection{Baseline methods}

This article constructs or chooses nine baseline models as follows, evaluating the pros and cons of the AIProbS compared with its predecessors of both classical and machine learning-based baselines.

Classical baselines include two models. As the bedrock, the ProbS \cite{zhang2007recommendation} is a necessary baseline to evaluate the improvement of the AIProbS. In addition, one might expect to base the recommendation directly on the nodal representations generated by the methods proposed in Sec.~\ref{generate_node_features}, not the ProbS framework. To test this strategy, this article constructs the Pure-DHC model, used to perform the recommendation by Eq.~(\ref{user-item_similarity_matrix}) based on the user-item similarity of their H-indices (\textit{i.e.}, nodal representations).

Machine learning-based baselines include seven models. To avoid the baseline pitfalls that have plagued earlier research on the comprehensive and objective evaluation of proposed models, this article further chooses seven representative machine learning-based models as baselines, among which were based on six different techniques of machine learning frameworks, including NeuMF \cite{he2017neural} based on deep neural networks, ConvNCF \cite{he2018outer} and SpectralCF \cite{zheng2018spectral} based on convolution operations, GCMC \cite{berg2017graph} based on graph auto-encoder frameworks, LINE \cite{tang2015line} based on random walking, NGCF \cite{wang2019neural} based on graph neural networks, and DGCF \cite{wang2020disentangled} based on attention mechanisms.

To guarantee the fairness and reproducibility of experiments, the implementation and evaluation of models were hosted to the RecBole \cite{zhao2021recbole}, a public open pipeline of recommender systems. One can get the codes used in this article through the hyperlink address posted in Sec.~\ref{data_and_code_available}.

\section{Results}
\label{Experimental_results}

Based on the experimental settings in Sec.~\ref{performance_evaluation}, this section presents the experimental results on the precision and robustness of the AIProbS and baseline models in Sec.~\ref{Precision analysis} and Sec.~\ref{Robustness analysis}, respectively, revealing their pros and cons in different recommendation scenarios.


\subsection{Precision analysis}
\label{Precision analysis}

As shown in Tabs.~\ref{LastFM}, \ref{100K} and \ref{1M}, the results on model precision are presented, where the length $N$ of the recommendation list is set to $10$, and each model's precision is averaged from its independently implementation based on $30$ different realizations. The values in parentheses indicate the percentage of improvement or decline in model precision of the AIProb model compared to the respective baseline models on each specific data set and evaluation metric, where the percentage of improvement is bold.

Conceivably, when speaking of the necessity of determinable and interpretable network representation and their utilization in link prediction, one might cast it into doubt: do not nodal representations generated by the two methods in Sec.~\ref{generate_node_features} be sufficient for link prediction? Can the precision of classical link prediction frameworks really be enhanced by being involved with nodal representations as intelligence? Are the machine learning-based network representation methods not precise enough? As shown in Tabs.~\ref{LastFM}, \ref{100K} and \ref{1M}, on all three data sets the Pure-DHC which directly utilizes the nodal representations generated by the methods in Sec.~\ref{generate_node_features} for recommendation achieved the worst model precision among the AIProbS and baseline models. That is to say, such generated nodal representations could be nothing with the recommendation if not be utilized in the ProbS or any other recommendation framework. After utilizing these nodal representations in the ProbS framework, as shown in Tabs.~\ref{LastFM}, \ref{100K} and \ref{1M}, the AIProbS outperformed the classical ProbS on all three data sets. When compared to machine learning-based baselines, the AIProbS indeed performed worse than some machine learning models, most obviously on MovieLens 1M. But it still can achieve state-of-the-art performance on model precision on LastFM, suggesting that nodal representations generated by the methods in Sec.~\ref{generate_node_features} may be able to abstract the underlying attributes of a complex network better than those learned by machine learning methods.

\begin{table}[ht]
\caption{\textbf{Results of model precision on LastFM.}}
\label{LastFM}
\begin{indented}
\item[]\begin{tabular}{@{}cccc@{}}
\br
\multirow{2}{*}{Models} & \multicolumn{3}{c}{LastFM}                                                                                                  \\ \cmidrule(r){2-4}
                        & Recall@10                                 & MRR@10                                    & NDCG@10                                   \\ \mr
Pure-DHC                & 0.004                                  & 0.006                               & 0.003                                     \\
ProbS                   & 0.170 (\textbf{+7.5}\%)                & 0.308 (\textbf{+9.6}\%)             & 0.166(\textbf{+9.2}\%)                   \\
AIProbS                 & \textbf{0.184}                         & \textbf{0.340}                      & \textbf{0.183}                           \\ \midrule
NeuMF                   & 0.060 (\textbf{+67.5}\%)               & 0.092 (\textbf{+73.0}\%)            & 0.050 (\textbf{+72.5}\%)                   \\
ConvNCF                 & 0.056 (\textbf{+69.7}\%)               & 0.090 (\textbf{+73.6}\%)            & 0.048 (\textbf{+73.8}\%)                   \\
SpectralCF              & 0.066 (\textbf{+63.9}\%)               & 0.120 (\textbf{+64.7}\%)            & 0.062 (\textbf{+66.1}\%)                           \\
GCMC                    & 0.121 (\textbf{+34.1}\%)               & 0.214 (\textbf{+37.0}\%)            & 0.116 (\textbf{+36.5}\%)                            \\
LINE                    & 0.149 (\textbf{+19.1}\%)               & 0.272 (\textbf{+20.0}\%)            & 0.145 (\textbf{+20.5}\%)                        \\
NGCF                    & 0.169 (\textbf{+8.2}\%)                & 0.301 (\textbf{+11.4}\%)            & 0.163 (\textbf{+11.0}\%)                            \\
DGCF                    & 0.177 (\textbf{+3.7}\%)                & 0.316 (\textbf{+7.1}\%)             & 0.172 (\textbf{+6.0}\%)                            \\ \br
\end{tabular}
\end{indented}
\end{table}

\begin{table}[ht]
\caption{\textbf{Results of model precision on MovieLens 100K.}}
\label{100K}
\begin{indented}
\item[]\begin{tabular}{@{}cccc@{}}
\br
\multirow{2}{*}{Models} & \multicolumn{3}{c}{MovieLens 100K}                                                                                                  \\ \cmidrule(r){2-4}
                        & Recall@10                                 & MRR@10                                    & NDCG@10                                   \\ \mr
Pure-DHC                & 0.017                                  & 0.084                               & 0.041                                      \\
ProbS                   & 0.208 (\textbf{+3.1}\%)                & 0.413 (\textbf{+4.8}\%)             & 0.236 (\textbf{+4.5}\%)                   \\
AIProbS                 & \textbf{0.215}                         & \textbf{0.434}                      & \textbf{0.248}                           \\ \midrule
NeuMF                   & 0.070 (\textbf{+67.2}\%)               & 0.187 (\textbf{+56.9}\%)            & 0.093 (\textbf{+62.3}\%)                   \\
ConvNCF                 & 0.099 (\textbf{+53.9}\%)               & 0.245 (\textbf{+43.5}\%)            & 0.125 (\textbf{+49.5}\%)                   \\
SpectralCF              & 0.124 (\textbf{+42.4}\%)               & 0.293 (\textbf{+32.5}\%)            & 0.153 (\textbf{+38.2}\%)                           \\
GCMC                    & 0.196 (\textbf{+8.9}\%)                & 0.400 (\textbf{+7.9}\%)             & 0.232 (\textbf{+6.4}\%)                            \\
LINE                    & 0.190 (\textbf{+11.3}\%)               & 0.391 (\textbf{+9.8}\%)             & 0.225 (\textbf{+9.0}\%)                         \\
NGCF                    & 0.245 (-14.3\%)                        & 0.481 (-10.8\%)                     & 0.293 (-18.4\%)                             \\
DGCF                    & 0.236 (-9.9\%)                         & 0.458 (-5.5\%)                      & 0.278 (-12.3\%)                            \\ \br
\end{tabular}
\end{indented}
\end{table}

\begin{table}[ht]
\caption{\textbf{Results of model precision on MovieLens 1M.}}
\label{1M}
\begin{indented}
\item[]\begin{tabular}{@{}cccc@{}}
\br
\multirow{2}{*}{Models} & \multicolumn{3}{c}{MovieLens 1M}  \\ \cmidrule(r){2-4}
                        & Recall@10                                 & MRR@10                                    & NDCG@10                                   \\ \mr
Pure-DHC                & 0.002                                     & 0.031                                     & 0.014                                     \\
ProbS                   & 0.108 (\textbf{+17.6\%})                  & 0.352 (\textbf{+15.1\%})                  & 0.177 (\textbf{+15.9\%})                  \\
AIProbS                 & \textbf{0.131}                            & \textbf{0.414}                            & \textbf{0.210\%}                          \\ \midrule
NeuMF                   & 0.032 (\textbf{+75.4\%})                  & 0.128 (\textbf{+69.1\%})                  & 0.053 (\textbf{+74.6\%})                  \\
ConvNCF                 & 0.073 (\textbf{+44.2\%})                  & 0.255 (\textbf{+38.4\%})                  & 0.128 (\textbf{+38.9\%})                  \\
SpectralCF              & 0.147 (-11.5\%)                           & 0.416 (-0.5\%)                            & 0.236 (-12.2\%)                           \\
GCMC                    & 0.152 (-15.4\%)                           & 0.421 (-1.7\%)                            & 0.240 (-14.4\%)                           \\
LINE                    & 0.153 (-16.6\%)                           & 0.423 (-2.1\%)                            & 0.236 (-12.3\%)                           \\
NGCF                    & 0.162 (-23.4\%)                           & 0.442 (-6.7\%)                            & 0.254 (-21.0\%)                           \\
DGCF                    & 0.172 (-31.0\%)                           & 0.460 (-11.1\%)                           & 0.266 (-26.8\%)                           \\ \br
\end{tabular}
\end{indented}
\end{table}

To put these results in more general terms, it is definite that designing control experiments to guarantee the comprehensiveness and objectivity of model performance evaluation is indispensable because, as shown in Tabs.~\ref{LastFM}, \ref{100K} and \ref{1M}, the comparative predominance between different models or even that between the classical and machine learning-based models are distinctive. For instance, compared to its predecessor (the ProbS framework), the AIProbS at best improved the Recall@$10$ by $17.6\%$ on MovieLens 1M and at worst, by $3.1\%$ on MovieLens 100K. Such a $14.5\%$ gap shows that the predominance of the AIProbS over the ProbS is not necessarily that significant in all recommendation scenarios. Overall, on MovieLens 1M, although it achieved an appreciable improvement over the ProbS, the AIProbS still performed worse than the other five machine learning-based models, revealing the predominance of the machine learning-based frameworks over the classical ones on this data set. However, that predominance faded on MovieLens 100K because only two machine learning-based models (\textit{i.e.}, NGCF and DGCF) outperformed the AIProbS. On LastFM, none of the machine learning-based models outperformed the AIProbS, in other words, but the AIProbS achieved state-of-the-art performance on model precision.

Figuring out the determinant factors of model precision in different recommendation scenarios is not easy and intuitive, not to mention accurately predicting a model's performance for one specific scenario. Still, on the advice of the clues given in Tabs.~\ref{LastFM}, \ref{100K} and \ref{1M}, some discoveries could be summarized as follows. (1) The machine learning-based models might have a predominance on data sets with large scales. The recommendation scenario of MovieLens 1M and MovieLens 100K being equal, the machine learning-based models would show a more obvious predominance on the former with a comparative larger data scale than the latter. However, it is hard to assert that the distinctions of ratios of $|U|$ to $|V|$ and the sparsity of the two data sets play a silent role. (2) The classical frameworks might play a large role in the improvement after integration in recommendation scenarios with high sparsity. Since the sparsity of LastFM is the highest among the three data sets, where the machine learning-based models face fundamental limits on lack of enough user-item interactions for training, the AIProbS or the ProbS combined with or of the classical frameworks showed their predominance as a result of their network structure-oriented resolution. Nevertheless, the ratio of $|U|$ to $|V|$ of LastFM, which seems to be a little higher than the other two, could also be a decisive factor.

\subsection{Robustness analysis}
\label{Robustness analysis}

As revealed in Sec.~\ref{Precision analysis}, with the increase in data scale the precision of the AIProbS decreased compared with that of machine learning-based baselines, for the mechanism of data fitting (or pattern representation) adopted by machine learning methods can give fully to its play more suitably in scenarios with larger data scale. Nevertheless, it does not mean that the AIProbS is useless in those scenarios. Since hyperparameters having no realistic meanings could make a model's implementing process indeterminable and vague, machine learning-based models lost almost all the interpretability for results, like why a machine learning-based model generates some recommendations for a user. This problem is not confined to recommender systems but still haunts other applications requiring high interpretability for results, such as machine translation or knowledge graph completion. Besides, since tedious hyperparameter tuning is required for up to the optimal performance of a model, generally machine learning methods reach higher precision by sacrificing the model's computing efficiency. Such a strategy brings about heavy financial (\textit{i.e.}, computing resources) and time costs for the implementation on a huge data. In contrast, the AIProbS is determinable and its results are interpreted, meaning that this model could be more suitable for applications requiring high interpretability and for scenarios with huge data scales but insufficient computing resources.

\begin{figure}[ht]
\includegraphics[scale=0.46]{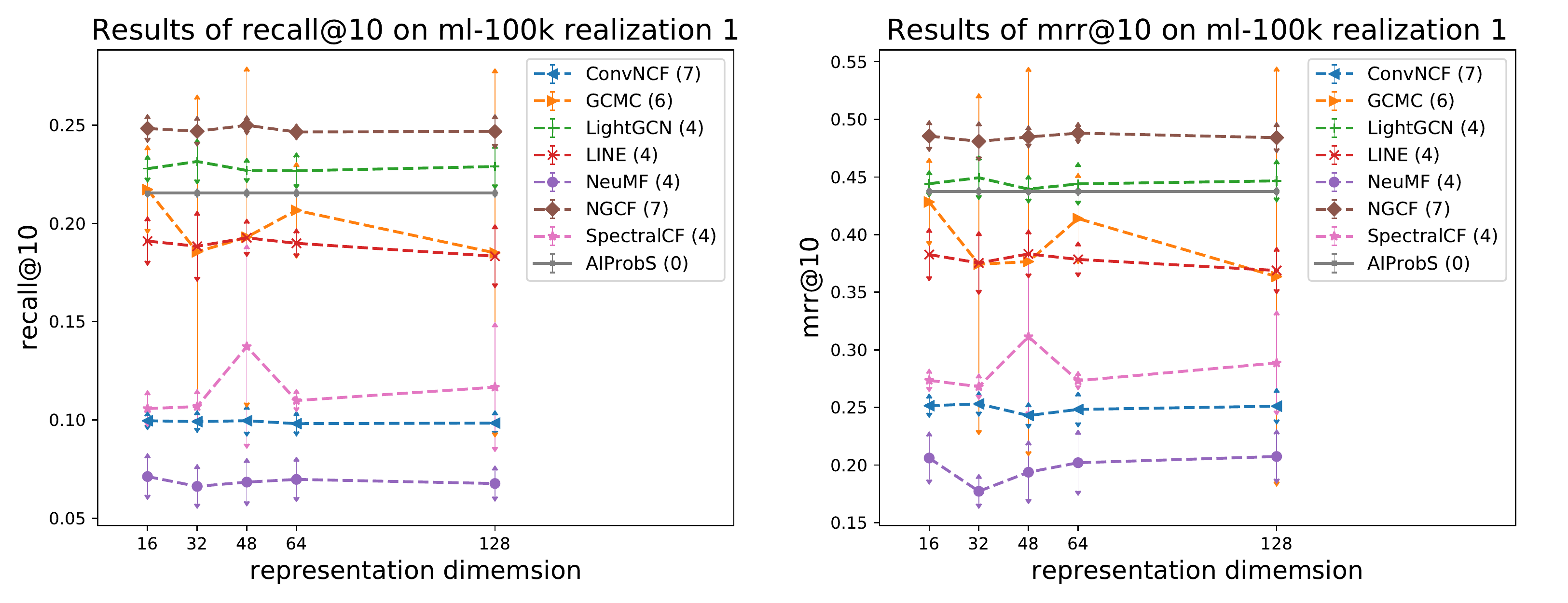}
\caption{\textbf{Relation between model precision and representation dimension reflected on ml-100k realization 1}}
\label{P1}
\end{figure}

\begin{figure}[ht]
\includegraphics[scale=0.46]{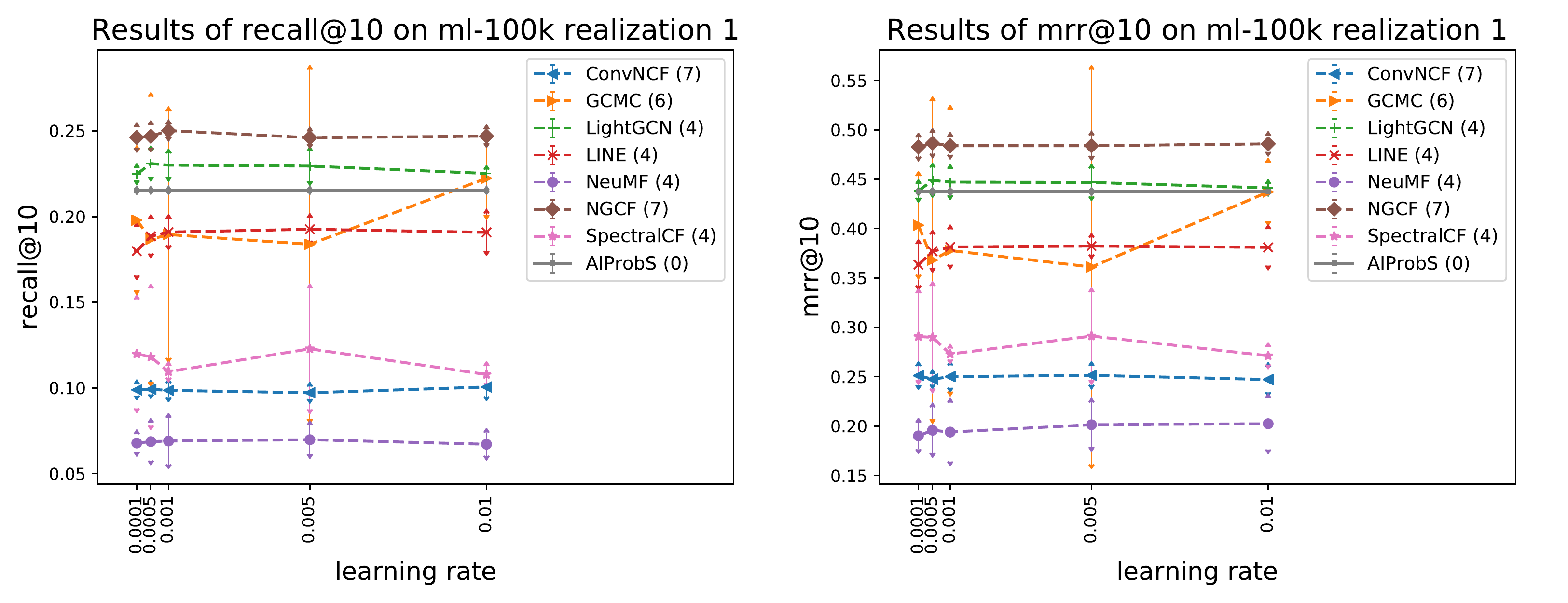}
\caption{\textbf{Relation between model precision and learning rate reflected on ml-100k realization 1}}
\label{P2}
\end{figure}

\begin{figure}[ht]
\includegraphics[scale=0.46]{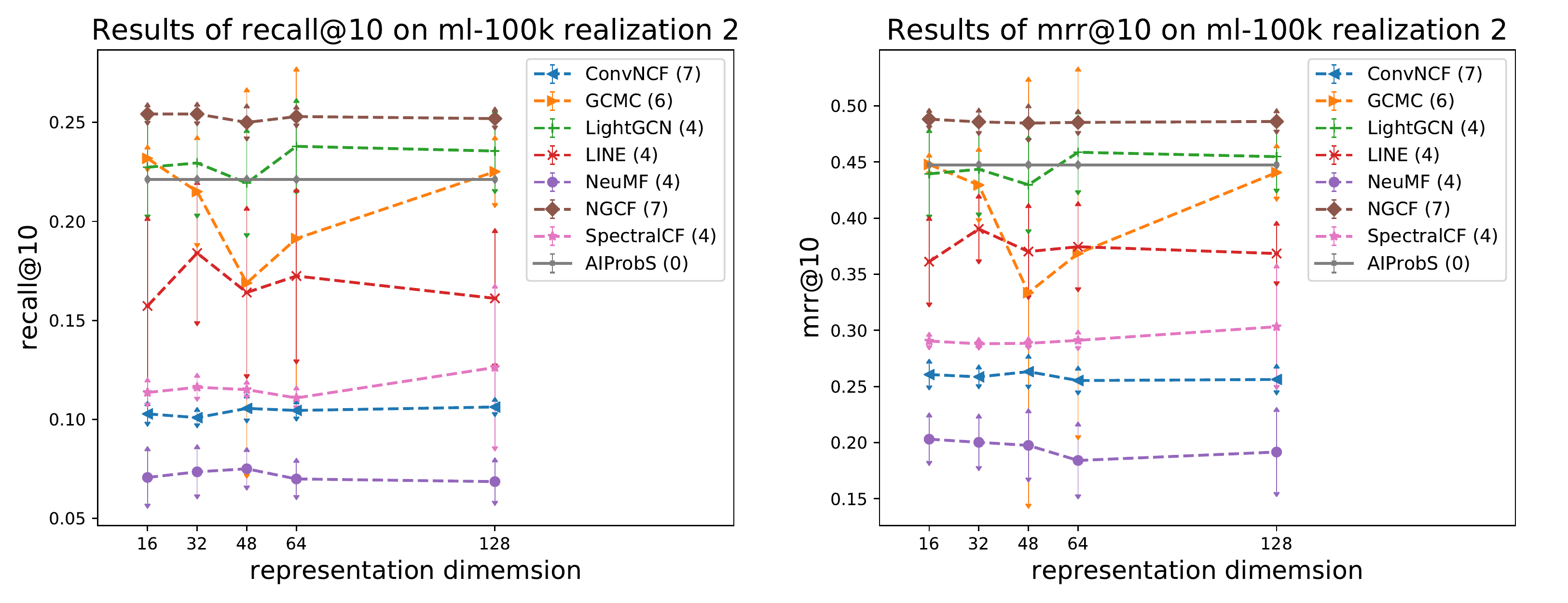}
\caption{\textbf{Relation between model precision and representation dimension reflected on ml-100k realization 2}}
\label{P3}
\end{figure}

\begin{figure}[ht]
\includegraphics[scale=0.46]{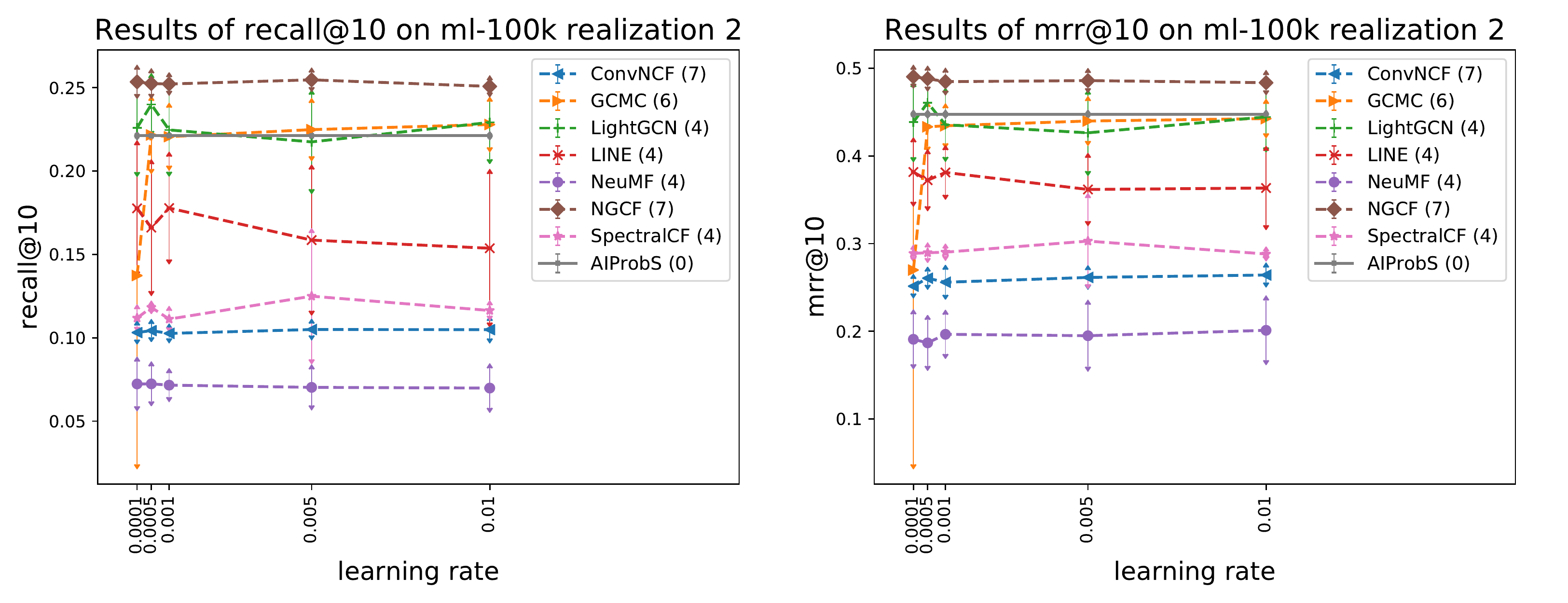}
\caption{\textbf{Relation between model precision and learning rate reflected on ml-100k realization 2}}
\label{P4}
\end{figure}

On two realizations of ml-100k for instance, Tabs.~\ref{P1}, \ref{P2}, \ref{P3}, and \ref{P4} present the relations between the different settings of two representative hyperparameters (\textit{i.e.}, representation dimension and learning rating) and a model's average precision when one hyperparameter is fixed and others are traversed within a specified search range, where the standard deviation of precision is presented by error arrow at a data point and each model's number of hyperparameters is presented in parentheses, reflecting a model's magnitude of performance fluctuation associated with setting disturbance, which actually can quantify the model's robustness. As shown in Tab.~\ref{P1}, the AIProbS had a stable performance on recall@10, since the representation dimension of its results is determined. However, as a hyperparameter different settings of representation dimension can largely influence machine learning-based baselines' precision. Although the performance on recall@10 of ConvNCF, LightGCN, and NGCF of different representation dimensions was relatively stable among machine learning-based baselines, that of GCMC and SpectralCF largely fluctuated with the change of representation dimension. For example, as for SpectralCF when the representation dimension is set to $48$ its average precision could be around $27\%$ higher than that when being set to $16$. On top of that, even when the representation dimension of SpectralCF is set to $48$, seemingly the optimal choice, its performance on recall@10 still faces a $125\%$ gap between the peaks of performance, flowed from the different settings of other hyperparameters. Similar fluctuations on machine learning-based baselines' precision were revealed by the influence of different settings of representation dimension on mrr@10 and by results shown in Tab.~\ref{P2} when considering learning rate as the controlled hyperparameter. Attributed to such the indeterminable performance of machine learning methods, one may have to repeatedly try different hyperparameter settings for a machine learning-based model to search out the optimal one, which is definitely computing resources-consuming and time costly. If an insufficient searching process turns out improper hyperparameter settings, the model could even end up with its worst performance. In contrast, once implemented the AIProbS can reach its optimal performance.

It may be said, though, that virtually determining a machine learning-based model's optimal hyperparameter settings is burdensome but can be once and for all based on one data set. However, it is not a fact. Similar experiments were done on another realization of ml-100k and the results presented in Tabs.~\ref{P3} and \ref{P4} revealed that the optimal hyperparameter settings of a machine learning-based model would be changed with the change of data set. For example, as shown in Tab.\ref{P1} the optimal representation dimension of SpectralCF was $48$ on ml-100k realization 1 but that on ml-100k realization 2 was changed to be $128$, as shown in Tab.~\ref{P3}. As a result, the searching process for the optimal hyperparameter settings of a machine learning-based model on a new data set appears to be inevitable. In contrast, with the change of data set the representation dimension of the AIProbS is still automatically determined, once implemented.

All in all, in the sense that the AIProbS provides a good trade-off with machine learning-based models on precision, interpretability, and determinacy.

\section{Discussion}

This article proposes two determinable and interpretable node representation methods. Different from other attempts like automated machine learning methods by computer scientists and computational theory by mathematics to search out and analyze the sub-optimal (or optimal) representation dimension of a machine learning-based network representation model, respectively, and to interpret the implementing process and the results come out of the model, from a perspective of physics the proposed two methods can substantially generate nodal representations with a determined dimension and interpretable elements, reaching its optimal performance once implemented. After utilizing these representations in link prediction for bipartite networks, experimental results showed that the AIProbS can make a good trade-off with machine learning-based models on precision, determinacy, and interpretability, indicating the effectiveness of nodal representations generated by the proposed two representation methods.

Importantly, these methods with good generalization may motivate further research. For example, nodal representations generated by the proposed two methods can also be utilized in machine learning-based models as initial features or network representation, and the AIProbS provides a unified architecture that various nodal representations generated by other methods, like machine learning-based methods, can be involved, which may further improve the precision of link prediction.

Nevertheless, like any model under the effect of the no-free-lunch theorem \cite{adam2019no} that no model can always perform well enough as expected in all different scenarios, the AIProbS has its disadvantages in some scenarios. With the increase in data scales, the AIProbS overall underperformed machine learning-based models on precision. Although the AIProbS can make a good trade-off with machine learning methods on precision and interpretability, in some applications where results' interpretability is unnecessary, like computer vision, machine learning methods seem like a better choice. Besides, since quantum machine learning is usually claimed as the next generation of machine learning, which can exponentially uplift a model's computing efficiency, would costly hyperparameter tuning be no longer an apprehension in the future? In other words, would determinable network representation that could sacrifice some precision but not representation learning-based (\textit{i.e.}, machine learning-based) methods that are adept in precision still be worthy of quantum computing devices in the future?

\section{Acknowledgments}

The author would like to acknowledge Linyuan L\"u, Hao Wang, and Fang Zhou for their discussions and suggestions.

\section{Data availability statement}
\label{data_and_code_available}

Data and codes are available at \url{https://github.com/pitteryue/AIProbS}.



\section*{References}
\bibliographystyle{iopart-num}
\bibliography{sample-base}



\clearpage
\begin{appendix}
\section{Proof of theorem 2.2}

\subsection{Preliminaries}

Before proving theorem 2.2, essential preliminaries are presented as follow.

\textbf{Definition A.1.} \textit{A simple matrix is defined as a matrix $A^{n \times n}$ when the algebraic and geometrical multiplicities of each of its eigenvalues are equal.}

\textbf{Definition A.2.} \textit{A normal matrix is defined as a matrix $A^{n \times n}$ if and only if $AA^T=A^TA$ when $A^{n \times n}$ is a real matrix or $AA^H=A^HA$ when $A^{n \times n}$ is a complex matrix.}

\textbf{Definition A.3.} \textit{An idempotent matrix is defined as a matrix $A^{n \times n}$ if and only if $A^2=A$.}

\textbf{Lemma A.1.}\label{lemma_1} \textit{Let $A$ be a $n \times n$ simple matrix with eigenvalues $\lambda_1, \lambda_2, ..., \lambda_n$. Then $A$ can be diagonalized by}
\begin{equation}
P^{-1} A P =  \mbox{diag}(\lambda_1, \lambda_2, ..., \lambda_{n}),
\end{equation}
\textit{where $P$ is an invertible matrix.}

\textit{Proof.} For any matrix $A$, there exists an invertible matrix $P$ such that $\displaystyle P^{-1}AP = J = \mbox{diag}(J_1(\lambda_1), J_2(\lambda_2), ..., J_r(\lambda_r))$, where
\begin{equation}
J_i(\lambda_i)=\left(
\begin{array}{cccc}
 \lambda_i   & 1     &           &   \\
            &...    &...        &   \\
            &       & \lambda_i &1  \\
            &       &           &\lambda_i
\end{array} \right )
\end{equation}
is called a Jordan block, and the matrix $J$ is called the Jordan standard form of $A$.

The number of Jordan blocks corresponding to the same eigenvalue is the geometric multiplicity of that eigenvalue. The sum of the orders of all Jordan blocks corresponding to the same eigenvalue is the algebraic multiplicity of that eigenvalue

By the definition of a simple matrix, where the algebraic and geometrical multiplicities of each of its eigenvalues are equal, the Jordan blocks of a simple matrix $A^{n \times n}$ can be denoted by $\displaystyle J_i(\lambda_i) = (\lambda_i)$, a $1 \times 1$ matrix. Then, for a simple matrix $A^{n \times n}$, we have:
\begin{equation}
P^{-1}AP = \left(
\begin{array}{cccc}
J_i(\lambda_1)  &                   &       & \\
                & J_2(\lambda_2)    &       & \\
                &                   &...    & \\
                &                   &       & J_n(\lambda_n) \\
\end{array}\right)
=\left(
\begin{array}{cccc}
\lambda_1  &                   &       & \\
                & \lambda_2    &       & \\
                &                   &...    & \\
                &                   &       & \lambda_n \\
\end{array}\right).
\end{equation} \hspace{15.3cm} $\square$

\subsection{Proof of Theorem 2.2}
Apparently, as a symmetric matrix, the adjacency matrix $B^{(m+n) \times (m+n)}$ is a normal matrix, so it is a simple matrix. Then, by Lemma~\ref{lemma_1}, we have:
\begin{equation}
\label{B}
B = P \mbox{diag}(\lambda_1, \lambda_2, ..., \lambda_{m+n})P^{-1},
\end{equation}
where $\lambda_{1}, \lambda_{2},...,\lambda_{m+n}$ are the $m+n$ eigenvalues of $B$.

Given $P$ is an invertible matrix, we can denote
\begin{equation}
\label{P}
P=(\nu_1, \nu_2, ..., \nu_{m+n})
\end{equation}
consisting of linearly independent column vectors $\nu_i$, where $B \nu_i = \lambda_i \nu_i (i=1,2,...,m+n)$. Since $\displaystyle BP = P \mbox{diag}(\lambda_1, \lambda_2, ..., \lambda_{m+n})$, we have $\displaystyle B=P \mbox{diag}(\lambda_1, \lambda_2, ..., \lambda_{m+n}) P^{-1}$. Then:

\begin{eqnarray}
\label{combine_eq_1}
B^T &=(P^{-1})^T \mbox{diag}(\lambda_1, \lambda_2, ..., \lambda_{m+n}) P^T \\
  & =  (P^T)^{-1} \mbox{diag}(\lambda_1, \lambda_2, ..., \lambda_{m+n}) P^T.
\end{eqnarray}

Let $\displaystyle \omega_1, \omega_2, ..., \omega_{m+n}$ be the $m+n$ eigenvectors of $B^T$, which are linearly independent column vectors, we have $\displaystyle B^T(\omega_1, \omega_2, ..., \omega_{m+n}) = (\omega_1, \omega_2, ..., \omega_{m+n}) \mbox{diag}(\lambda_1, \lambda_2, ..., \lambda_{m+n})$. Then:
\begin{equation}
\label{combine_eq_2}
B^T = (\omega_1, \omega_2, ..., \omega_{m+n}) \mbox{diag}(\lambda_1, \lambda_2, ..., \lambda_{m+n})(\omega_1, \omega_2, ..., \omega_{m+n})^{-1}.
\end{equation}

Apparently, combining Eq.~(\ref{combine_eq_1}) and Eq.~(\ref{combine_eq_2}) we get $(P^{T})^{-1} = (\omega_1, \omega_2, ..., \omega_{m+n})$, so we can denote $P^{-1} =((P^{-1})^T)^T = (((P^{T})^{-1})^T  =  (\omega_1, \omega_2, ..., \omega_{m+n})^T$. That is:
\begin{equation}
\label{P-1}
P^{-1}=\left(
\begin{array}{cccc}
\omega_1^T\\ \omega_2^T\\ ... \\ \omega_{m+n}^T
\end{array}\right).
\end{equation}

Based on Eqs.~(\ref{B}), (\ref{P}) and (\ref{P-1}), we can decompose $B$ by
\begin{eqnarray}
B&=\left(
\begin{array}{c}
\nu_1, \nu_2,...,\nu_{m+n}
\end{array}\right)
\left(\begin{array}{cccc}
\lambda_1  &0           &...    &0   \\
0          &\lambda_2   &...    &0   \\
...        &...         &...    &... \\
0          &0           &...    &\lambda_{m+n}
\end{array}\right)
\left(\begin{array}{cccc}
\omega_1^T\\ \omega_2^T\\ ... \\ \omega_{m+n}^T
\end{array}\right)
\\&=\sum_{i=1}^{m+n}\lambda_i \nu_i \omega_i^T.
\end{eqnarray}

Furthermore, let $\displaystyle B_i=\nu_i \omega_i^T$. Since $\displaystyle P^{-1} P = E_{m+n}$ (\textit{i.e.}, $\displaystyle \omega_i^T \nu_j = \left \{ \begin{array}{cc}
1 & j=i, \\
0 & j \neq i,
\end{array} \right.$), $B_i(i=1,2,...,m+n)$ are idempotent matrices satisfying $\displaystyle B_iB_j = \left \{ \begin{array}{cc}
B_i & j=i, \\
0 & j \neq i,
\end{array} \right.$.

Finally, we come to the theorem that $B$ can be decomposed by
\begin{equation}
B = \sum_{i=1}^{m+n}\lambda_i B_i,
\end{equation}
where $\lambda_i$ is the i-th eigenvalue of $\displaystyle B^{(m+n)\times (m+n)}$ and $\displaystyle B_i$ is the corresponding idempotent matrix. \hspace{13.8cm}$\square$

\clearpage
\section{Proof of theorem 2.3}

\textit{Proof.} Since $\displaystyle R = A \cdot (D_{I}  \circ  A)^T \cdot (D_{U}  \circ  A) \Leftrightarrow R^T = (D_U \circ A)^T \cdot (D_I \circ A) \cdot A^T$ holds, here let the operator $T'=(D_U \circ A)^T \cdot (D_I \circ A)$.

In the first place, construct a non-empty complete metric space $(\mathbb{R}^{n \times m},d_{\max})$ based on $\mathbb{R}^{n \times m}$. Since any two norms on a finite-dimensional linear space are equivalent, for simplicity we choose the norm-induced metric $\displaystyle d_{\max}(A,B)=\max_{i,j}\{|a_{ij}|\}, 1 \leq i \leq m, 1 \leq j \leq n$. It is easy to prove that $(\mathbb{R}^{n \times m}, d_{\max})$ is a non-empty complete metric space, as follows.

There is no doubt that $(\mathbb{R}^{n \times m}, d_{\max})$ is a metric space. Furthermore, assume $\{X\}$ is a Cauchy sequence in $\mathbb{R}^{n \times m}$, we have that $\forall \varepsilon>0, \exists N \in \mathbb{N}$, s.t., $m, n > N, d(X^{(n)}-X^{(m)})=\max_{i,j}\{|X_{ij}^{(n)}-X_{ij}^{(m)}|\}<\varepsilon$. For any sequence $\{X_{ij}\}$ with fixed $(i,j)$, we have that $\forall \varepsilon>0, \exists N \in \mathbb{N}$, s.t., $m, n > N, d(X_{ij}^{(n)}, X_{ij}^{(m)}) \leq \max_{i,j}\{|X_{ij}^{(n)}-X_{ij}^{(m)}|\}<\varepsilon$. So sequence $\{X_{ij}\}$ is a Cauchy sequence. Since $\mathbb{R}$ is complete, $X_{ij}^*$ exists such that $X_{ij}^{(n)} \longrightarrow X_{ij}^*$ when $n \longrightarrow \infty$, where $X_{ij}^* \in \mathbb{R}$. So $X^*$ exists such that $X^{(n)} \longrightarrow X^*$, when $n \longrightarrow \infty$. Apparently, $X^* \in \mathbb{R}^{n \times m}$. Therefore, $(\mathbb{R}^{n \times m},d_{\max})$ is a non-empty complete metric space.

Then, according to the Banach fixed point theorem on a non-empty complete metric space, the operator
\begin{eqnarray}
T'&=(D_U \circ A)^T \cdot (D_I \circ A) \\
  &= \left(
\begin{array}{cccc}
\frac{1}{K_{I_1}}\sum_{h=1}^{m}\frac{a_{h1}a_{h1}}{K_{U_h}} &\frac{1}{K_{I_2}}\sum_{h=1}^{m}\frac{a_{h1}a_{h2}}{K_{U_h}}
& ...
&\frac{1}{K_{I_n}}\sum_{h=1}^{m}\frac{a_{h1}a_{hn}}{K_{U_h}} \\
\frac{1}{K_{I_1}}\sum_{h=1}^{m}\frac{a_{h2}a_{h1}}{K_{U_h}}
&\frac{1}{K_{I_2}}\sum_{h=1}^{m}\frac{a_{h2}a_{h2}}{K_{U_h}}
& ...
&\frac{1}{K_{I_n}}\sum_{h=1}^{m}\frac{a_{h2}a_{hn}}{K_{U_h}} \\
... &... &... &...\\
\frac{1}{K_{I_1}}\sum_{h=1}^{m}\frac{a_{hn}a_{h1}}{K_{U_h}}
&\frac{1}{K_{I_2}}\sum_{h=1}^{m}\frac{a_{hn}a_{h2}}{K_{U_h}}
& ...
&\frac{1}{K_{I_n}}\sum_{h=1}^{m}\frac{a_{hn}a_{hn}}{K_{U_h}} \\
\end{array}\right),
\end{eqnarray}
where $a_{**} = 0$ or $1$ and $\displaystyle \frac{a_{**}a_{**}}{K_*}=0$ if $K_*=0$, is a contraction mapping on $(\mathbb{R}^{n \times m}, d_{\max})$. Finally, we have that $T$ is a contraction mapping on $(\mathbb{R}^{m \times n}, d_{\max})$, meaning that the iterative process $A \longleftarrow A \cdot T$ is convergent to a fixed point $A^*$.  \quad \quad \quad \quad \quad \quad \quad \quad \quad \quad $\square$

\clearpage
\section{Other combinations of the AIProbS}
\label{other_combinations}

Following Sec.~\ref{AIProbS}, the user-item similarity based on $F_U$ and $F_I$ can be calculated by other metrics, like covariance (Cov), dot product, Euclidean Distance (ED) and Pearson correlation coefficient (Pearson), as follows.

\subsection{Covariance (Cov)}
\begin{equation}
    S_{\mbox{Cov}}=\frac{1}{s} (F_U- \overline{F_U}) \cdot (F_I-\overline{F_I})^T,
\end{equation}
where $\displaystyle \overline{F_U}=(\alpha^T, \alpha^T, ..., \alpha^T)$, $\displaystyle \overline{F_I}=(\beta^T, \beta^T, ..., \beta^T)$, $\displaystyle \alpha = (\frac{1}{s} \sum_{j=1}^s {F_U}_{1j}, \frac{1}{s} \sum_{j=1}^s {F_U}_{2j}, ..., \\ \frac{1}{s} \sum_{j=1}^s {F_U}_{mj})$, $\displaystyle \beta = (\frac{1}{s} \sum_{j=1}^s {F_I}_{1j}, \frac{1}{s} \sum_{j=1}^s {F_I}_{2j}, ..., \frac{1}{s} \sum_{j=1}^s {F_I}_{nj})$.
\subsection{dot product}
\begin{equation}
    S_{\mbox{dot\_product}}=F_U \cdot F_I.
\end{equation}
\subsection{Euclidean distance (ED)}
\begin{equation}
    S_{\mbox{ED}}=S = \sqrt{|-2F_U \cdot F_I^T +\mathcal{F_U}+\mathcal{F_I}|},
\end{equation}
where $\displaystyle \mathcal{F_U}^{m \times n}=(\alpha^T, \alpha^T, ..., \alpha^T)$, $\displaystyle \mathcal{F_I}^{m \times n}=(\beta, \beta, ..., \beta)^T$, $\displaystyle \alpha = (\sqrt{\sum_{j=1}^s{F_U}_{1j}^2}, \sqrt{\sum_{j=1}^s{F_U}_{2j}^2}, ..., \\\sqrt{\sum_{j=1}^s{F_U}_{mj}^2})$, $\displaystyle \beta = (\sqrt{\sum_{j=1}^s{F_I}_{1j}^2},  \sqrt{\sum_{j=1}^s{F_I}_{2j}^2}, ..., \sqrt{\sum_{j=1}^s{F_I}_{nj}^2})$.
\subsection{Pearson correlation coefficient (Pearson)}
\begin{equation}
    S_{\mbox{Pearson}}=\frac{(F_U-\overline{F_U})(F_I-\overline{F_I})}{\alpha^T \beta},
\end{equation}
where $\displaystyle \alpha=\big(\sqrt{\sum_{j=1}^s ({F_U}_{1j}-\overline{F_U}_{1*}})^2, \sqrt{\sum_{j=1}^s ({F_U}_{2j}-\overline{F_U}_{2*}})^2, ...,  \sqrt{\sum_{j=1}^s ({F_U}_{mj}-\overline{F_U}_{m*}})^2 \big)$, $\displaystyle \beta=\big(\sqrt{\sum_{j=1}^s ({F_I}_{1j}-\overline{F_I}_{1*}})^2, \sqrt{\sum_{j=1}^s ({F_I}_{2j}-\overline{F_I}_{2*}})^2, ..., \sqrt{\sum_{j=1}^s ({F_I}_{nj}-\overline{F_I}_{n*}})^2 \big)$.

After combining these metrics with the max-min normalization (M-M) operation by Eq.~(\ref{max-min_normalization}) or the proportioning (P) operation by Eq.~(\ref{proportioning}), the AIProbS model of diverse versions can be constructed. Tab.~\ref{other_combinations_precision} presents the results on model precision of these new combinations, based on the evaluation settings designed in Sec.~\ref{performance_evaluation} and MovieLens 100K for an instance, among which the combination that \textit{cosine + M-M + P} used in this article achieves the best performance.

\begin{table}[h]
\caption{Results of model precision based on other combinations on MovieLens 100K, as an instance.}
\label{other_combinations_precision}
\begin{indented}
\item[]\begin{tabular}{@{}llll@{}}
\br
\multicolumn{1}{c}{Combinations}       & \multicolumn{1}{c}{Recall@10} & \multicolumn{1}{c}{MRR@10} & \multicolumn{1}{c}{NDCG@10} \\
\mr
\textbf{cosine + M-M + P (used)}       & \textbf{0.215}                & \textbf{0.434}             & \textbf{0.248}              \\
cosine + M-M                       & 0.167                         & 0.405                      & 0.217                       \\
cosine                                 & 0.167                         & 0.405                      & 0.217                       \\
Cov + M-M + P          & 0.132                         & 0.318                      & 0.166                       \\
Cov + M-M                          & 0.125                         & 0.303                      & 0.158                       \\
Cov                                    & 0.125                         & 0.303                      & 0.158                       \\
dot\_product + M-M + P & 0.180                         & 0.398                      & 0.218                       \\
dot\_product + M-M                 & 0.136                         & 0.343                      & 0.180                       \\
dot\_product                           & 0.136                         & 0.343                      & 0.180                       \\
ED + M-M + P           & 0.197                         & 0.411                      & 0.231                       \\
ED + M-M                           & 0.147                         & 0.366                      & 0.193                       \\
ED                                     & 0.147                         & 0.366                      & 0.193                       \\
Pearson + M-M + P      & 0.207                         & 0.419                      & 0.238                       \\
Pearson + M-M                      & 0.158                         & 0.387                      & 0.206                       \\
Pearson                                & 0.158                         & 0.387                      & 0.206                       \\
\br
\end{tabular}
\end{indented}
\end{table}

\clearpage
\section{The schematics and pseudocodes of AIProbS}

The pseudocodes of the AIProbS are displayed as follows.

\begin{table}[ht]
\small
\begin{adjustbox}{center}
\begin{tabular}{l}
\toprule
\textbf{Pseudocodes.} The AIProbS model \\ \hline
\textbf{Input:} implicit interactions between $m$ users and $n$ items, \\
                \quad \quad \quad the length $N$ of the recommendation list. \\
\textbf{Output:} user's top-$N$ recommendations.\\

\textbf{1\,\,:} \textbf{Construct Adjacency Matrix}: construct $A^{m \times n}$ of the input;  \\
\textbf{2\,\,:} \textbf{Generate Feature}: generate $F_U, \,F_I$ by methods in Sec.~\ref{generate_node_features}; \\
\textbf{3\,\,:} \textbf{Calculate Similarity:} calculate $S^{m \times n}$ by Eq.~(\ref{user-item_similarity_matrix});\\
\textbf{4\,\,:} \textbf{foreach} \textit{ row vector $(S\circ A)_{i*}$ in $(S \circ A)^{m \times n}$} \textbf{do}\\
\textbf{5\,\,:} \quad $\displaystyle \max$ = max($(S \circ A)_{i*}$), \,$\min$ = min($(S \circ A)_{i*}$); \\
\textbf{6\,\,:} \quad $\displaystyle (S \circ A)_{ij} \leftarrow \frac{(S \circ A)_{ij} - \min}{ \max - \min}, \,j = 1,2, ..., n$; \\
\textbf{7\,\,:} \quad $\displaystyle W_{U_{ij}} = \frac{1}{(S \circ A)_{ij}}\sum_{k=1}^{n} (S \circ A)_{ik},\, j=1,2,...,n$;\\
\textbf{8\,\,:} \textbf{foreach} \textit{column vector $(S\circ A)_{*j}$ in $(S \circ A)^{m \times n}$} \textbf{do}\\
\textbf{9\,\,:} \quad $\displaystyle \max$ = max($(S \circ A)_{*j}$), \,$\min$ = min($(S \circ A)_{*j}$); \\
\textbf{10:} \quad $\displaystyle (S \circ A)_{ij} \leftarrow \frac{(S \circ A)_{ij} - \min}{ \max - \min}, \,i = 1,2, ..., m$;\\
\textbf{11:} \quad $\displaystyle W_{I_{ij}} = \frac{1}{(S \circ A)_{ij}}\sum_{k=1}^{m} (S \circ A)_{kj},\, i=1,2,...,m$;\\
\textbf{12:} $\displaystyle R = A \cdot W_{I}^T \cdot W_{U}$;\\
\textbf{13:} \textbf{foreach} \textit{ row vector $R_{i*}$ in $R$} \textbf{do}\\
\textbf{14:} \quad record the column indices of $R_{i*}$'s top-$N$ elements into $L_i$;\\
\textbf{15:} \textbf{return} $L^{m \times N}$ as user's top-$N$ recommendations; \\
\bottomrule
\end{tabular}
\end{adjustbox}
\end{table}

\begin{figure}[ht]
\centering
\includegraphics[scale=0.45]{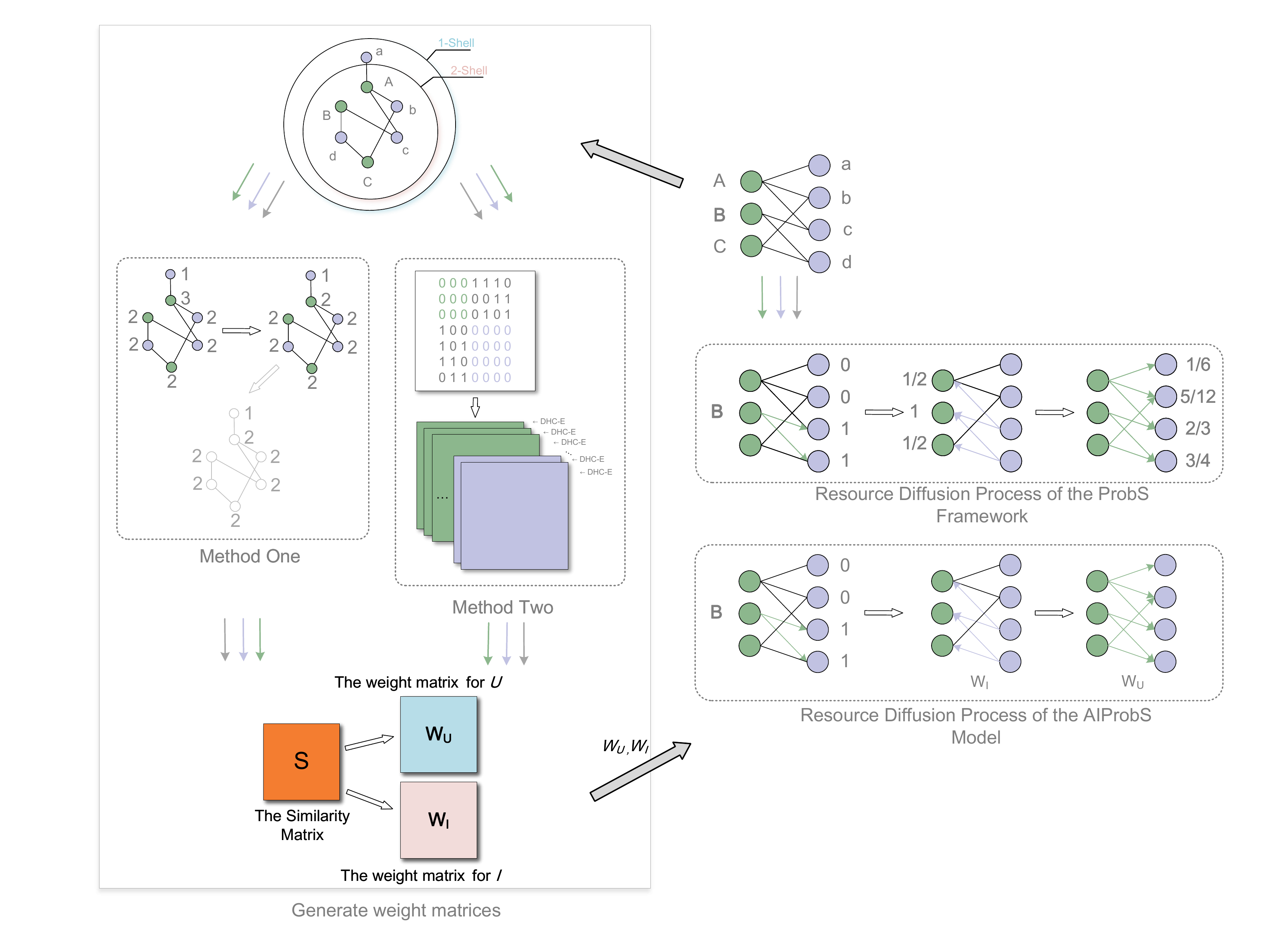}
\caption{\textbf{Schematics of the AIProbS model.} With a toy example, the schematics illuminate the processes of the two nodal feature generation methods and the recommendation based on the ProbS framework and the AIProbS model for Bob in a movie recommender system.}
\label{Schematics}
\end{figure}

\clearpage
Fig.~\ref{Schematics} illustrates the schematics of the AIProbS.

\end{appendix}

\end{document}